\documentclass[useAMS,usenatbib]{mn2e}
\usepackage{amsmath}
\usepackage{graphicx}
\usepackage{subfigure}
\usepackage{deluxetable}

\begin{document}

\title[Crossing the Threshold]{Evolving Molecular Cloud Structure and the Column Density Probability Distribution Function}
\author[Ward, Wadsley, \& Sills]{Rachel L. Ward, James Wadsley, and Alison Sills \\Department of Physics and Astronomy, McMaster University, Hamilton, ON, L8S 4M1, Canada}

\label{firstpage}
\maketitle

\begin{abstract} 
The structure of molecular clouds can be characterized with the probability distribution function (PDF) of the mass surface density.  In particular, the properties of the distribution can reveal the nature of the turbulence and star formation present inside the molecular cloud.  In this paper, we explore how these structural characteristics evolve with time and also how they relate to various cloud properties as measured from a sample of synthetic column density maps of molecular clouds.  We find that, as a cloud evolves, the peak of its column density PDF will shift to surface densities below the observational threshold for detection, resulting in an underlying lognormal distribution which has been effectively lost at late times.  Our results explain why certain observations of actively star-forming, dynamically older clouds, such as the Orion molecular cloud, do not appear to have any evidence of a lognormal distribution in their column density PDFs.  We also study the evolution of the slope and deviation point of the power-law tails for our sample of simulated clouds and show that both properties trend towards constant values, thus linking the column density structure of the molecular cloud to the surface density threshold for star formation.
\end{abstract}


\section{Introduction}\label{sec:Intro}

Giant molecular clouds (GMCs) are large regions of gas and dust, which undergo local gravitational collapse to form dense star-forming cores.  The morphology and structure of GMCs are significantly affected by the presence of turbulence which is highly supersonic on large-scales and leads to the formation of density enhancements such as filaments, clumps, and cores within which star formation occurs.  

The probability distribution function (PDF) of the mass density is one representation of the role of supersonic turbulence on the structure of molecular clouds.  Many numerical studies \citep[e.g.][]{vazquez1994,NP99,PN02,ostriker2001,KM05,HC08} have shown that a lognormal shape is expected for the density PDF of an isothermal, supersonic turbulent gas.  

Recent observations by \citet{kain2009} showed that column density PDFs for molecular clouds also exhibit a lognormal distribution.  However, they observed this distribution only in quiescent clouds which are not actively undergoing star formation.  The column density PDFs for star-forming molecular clouds, such as the Taurus Molecular Cloud, have an underlying lognormal shape with the addition of a power-law tail at high column densities.  This distribution is reminiscent of the stellar initial mass function (IMF) whose form is often approximated either by a lognormal \citep{chabrier} or by a power-law \citep{kroupa,salpeter} at the high-mass end of the spectrum.  These results suggest that characterising the shape of the density and column density PDF of molecular clouds is crucial for understanding the origin of the stellar IMF.  

Several numerical studies \citep{balle2011,knw11,tassis} have subsequently confirmed the presence of a power-law tail in the density and column density PDFs from three-dimensional simulations of molecular clouds collapsing to form stars.  These authors argue that the tail develops over time and its strength grows as more stars are formed and as gravity becomes dominant over turbulence.   

\citet{kain2011} explored the nature of the transition from a lognormal shape at low column densities to a power-law shape at higher column densities for actively star-forming clouds.  The authors showed that the changing shape of the PDF can be interpreted as the transition from a diffuse cloud dominated by turbulence to a clumpy gravitationally-dominated structure.  \citet{kain2011} showed that the `structural transition' of the PDF offers a physical explanation for the star formation threshold and for the correlation between the star formation rate and the mass of dense gas above the threshold in molecular clouds \citep{lada2010}.  Throughout this paper, we will refer to the threshold at which the structural transition occurs as $\Sigma_{\text{tail}}$ where all gas above this threshold is considered dense enough to form stars. 

There are several properties of the distribution which characterise the shape of the column density PDF for molecular clouds.  In this paper, we explore how these structural characteristics evolve with time and also how they relate to various cloud properties as measured from a sample of synthetic column density maps of bound and unbound molecular clouds.  Column density PDFs derived from molecular cloud observations can be ordered into an assumed sequence to demonstrate the evolutionary process from lognormal shape to the development of a power-law tail.  Simulators have shown this to be the case; however, the evolution of column density PDFs of simulated clouds in an observational context has not yet been considered.  Observational limits and thresholds naturally interfere with estimates of cloud properties.  In this paper, we show that thresholds can significantly affect the way we interpret molecular cloud structure and evolution, particularly for older, actively star-forming clouds like the Orion molecular cloud.

In Section~\ref{sec:methods}, we review the details of our simulations, the methods used to produce our synthetic column density maps, and the criteria for cloud selection first outlined in \citet{paper1}.  We describe our method of tracking clouds through time and present our findings on the evolution of their structural properties in Section~\ref{sec:results}.  Lastly, in Section~\ref{sec:discussion}, we summarize our results and discuss the implications of our conclusions. 

\section{Methods}\label{sec:methods}

We use simulations of molecular clouds to study the evolution of their structure and its effect on star formation.  Simulations allow us to explore the three-dimensional properties of the cloud in order to better understand the two-dimensional properties which are observed.  We created column density projections of our simulations to compare more directly to observations.  The dynamic properties of our synthetic clouds were the subject of a previous study by \citet{paper1} and full details of the simulations, column density maps, and the cloud selection process are provided in that paper.  What follows is a brief overview highlighting the key details of our method.

\begin{table}
	\caption{Initial Conditions for the Simulations \label{table:initcondit}}
	\centering
	\begin{tabular}{c c c c c c c c}
	\\
	Id &  Radius    & $\sigma_{3\text{D}}$ & n$_{\text{initial}}$ & t$_{ff}$ & $\alpha_{\text{initial}}$\\
		         & (pc) & (km s$^{-1}$) & (cm$^{-3}$) & (Myr) &  \\
	\hline\hline   
	9  & 30 & 2.93 & 7.69 & 22.4 & 2\\
	10 & 21.2 & 3.49 & 21.8 & 13.3 & 2\\
	11 & 15 & 4.15 & 61.5 & 7.9 & 2\\
	12 & 10.6 & 4.93 & 174 & 4.7 & 2\\
  13  & 30 & 2.07 & 7.69 & 22.4 & 1\\
  14 & 21.2 & 2.47 & 21.8 & 13.3 & 1\\
  15 & 15 & 2.93 & 61.5 & 7.9 & 1\\
  16 & 10.6 & 3.49 & 174 & 4.7 & 1\\
	\hline
  \end{tabular}
  \tablecomments{Reproduction of Table 1 from \citet{paper1}}
\end{table}

Our synthetic clouds were selected from initially bound 50 000 M$_{\odot}$ regions of the ISM, simulated using the smoothed particle hydrodynamics code \textsc{Gasoline} \citep{gasoline}.  Throughout this paper we only consider the cases where the initial global virial parameter of the volumes is less than or equal to 2 (or gravitationally bound) to avoid rapid dispersal due to high turbulent motions.  The initial conditions for these cases cover a range of initial densities and Mach numbers listed in Table 1 of \citet{paper1} for volumes labelled by IDs 9 - 16.  This table with the relevant cases has been reproduced for this work (Table 1).  Our particle mass is 8.85 $\times$ 10$^{-3}$ M$_{\odot}$ with a mass resolution limit for fragmentation of 0.6 M$_{\odot}$, resulting in simulations which are highly resolved and include only gravity and decaying turbulence with large-scale modes in non-periodic boxes.  We use the \citet{BB05} equation of state, whose opacity limit for fragmentation is greater than our maximum gas density, resulting in simulations which are entirely optically thin and isothermal at 10 K.  As a result, our column density projections are equivalent to synthetic observations derived from radiative transfer calculations in the limit of optically thin gas.  Detailed descriptions of the methods used to produce our synthetic column density maps can be found in \citet{paper1,ward2012}. Approximately 60 clouds at various stages of their evolution were selected from the column density maps of each turbulent region using surface density contours to identify the peak emission.  The selection process ensures that the clouds in our sample have column densities which exceed the lower limit for detection in dust extinction measurements \citep[A$_{\text{v}}$ $\sim$ 0.5;][]{beaumont2012,kain2009}.  This limit also approximates the threshold at which atomic hydrogen begins to self-shield against UV radiation ($\sim$ 10$^{20-21}$ cm$^{-2}$) leading to the formation of molecular hydrogen.  These synthetic clouds have sizes and surface densities which are analogous to observed molecular clouds, allowing for a more direct comparison to observations.  The spatial resolution of our column density maps is 6000 AU px$^{-1}$, which corresponds to an angular resolution of $\sim$ 45'' for a cloud at a distance of 140 pc, such as the Taurus molecular cloud \citep{nutter}.  

Although the simulation volumes are bound overall with global virial parameters $\leq$ 2, the clouds within the volume have local virial parameters ranging from 0.5 to 6.5 due to localised regions of collapse and turbulence.  Of the 58 clouds in our sample, over 30\% are gravitationally bound with local virial parameters $\leq$ 2.  This allows us to explore the structural evolution of both bound and unbound clouds, since the local virial parameters of the clouds remain steady and evolve minimally for the duration of the simulations.

We used sink particles to represent the high density regions of the simulations (inserted at densities greater than 10$^6$ cm$^{-3}$) based on the formation criteria of \citet{federrath10}.  Each sink corresponds to a star-forming core or a cluster-forming core with a radius of 500 AU.  We find that all of our clouds eventually form stars regardless of whether or not they are bound overall.  Since our simulations do not model radiative feedback, we do not include clouds in our study which have converted more than 35\% of their mass to stars.  

\section{Results}\label{sec:results}

In order to compare directly to observations, we only included the pixels from our maps with surface densities greater than $\sim$ 10 M$_{\odot}$ pc$^{-2}$.  This minimum surface density corresponds to the threshold for detection in extinction maps -- A$_{\text{v}}$ $\sim$ 0.5 \citep{kain2009} where the conversion factor is $\Sigma$/A$_{\text{v}}$ = 20 M$_{\odot}$ pc$^{-2}$ mag$^{-1}$ \citep{bohlin,lombardi2010,kain2011,lombardi2014}.  We also removed the sink particles from our column density maps, which is common practice for the case of embedded stars in studies of molecular clouds (in CO and extinction-based data) due to their tendency to bias measurements \citep{kain2009, goodman09, KFH13}.

\begin{figure*}
\begin{center}
   \includegraphics[scale=0.7,clip=true,trim=1cm 0cm 2cm 3cm]{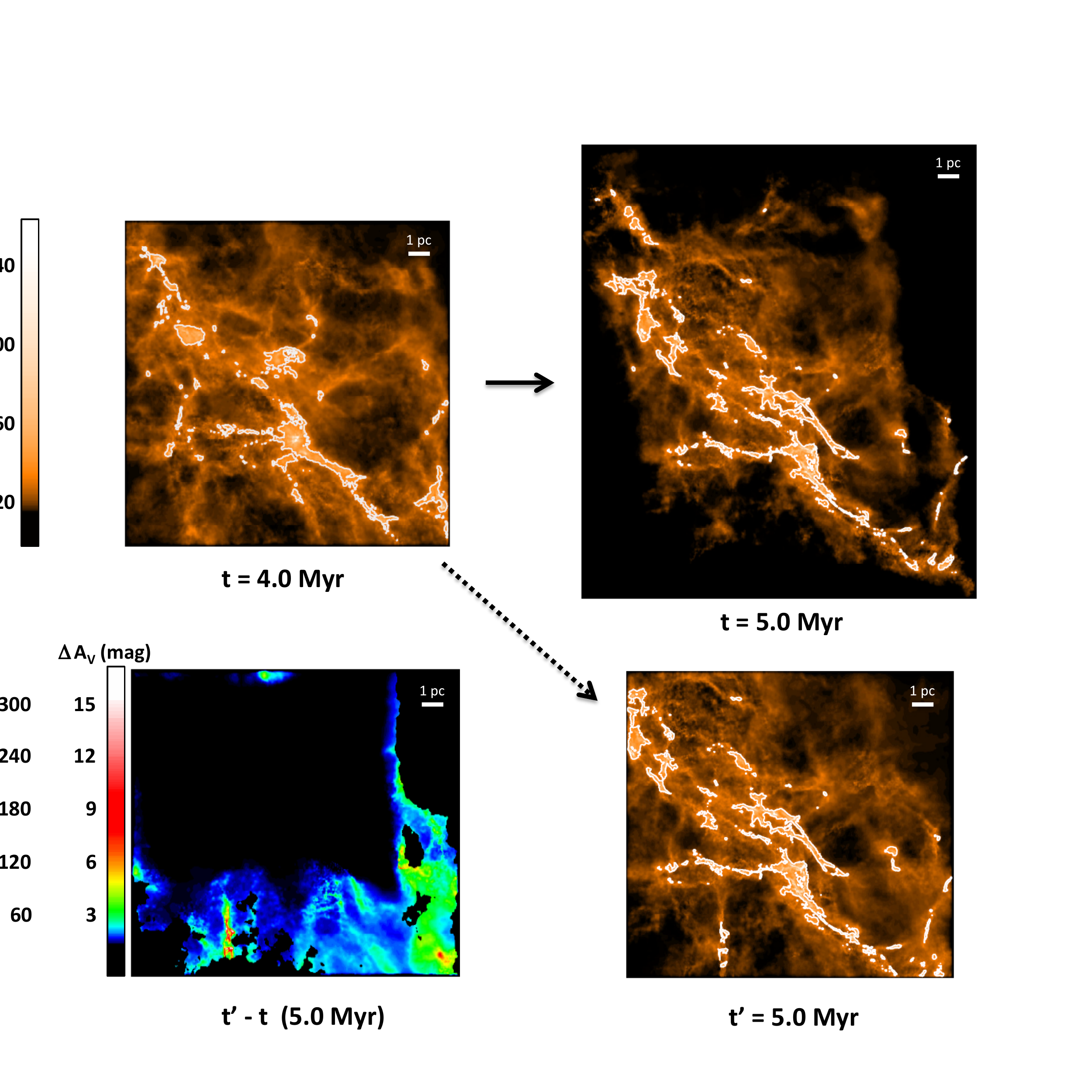}
   \caption{\label{fg:evolclouds_maps}
  Synthetic column density maps for the evolution of a single cloud.  The solid arrow shows the actual evolution of the particles from t=4.0 to t=5.0 Myr and the dotted arrow shows the perceived evolution of the cloud taken from independent snapshots at t=4.0 Myr and t'=5.0 Myr using an observationally-motivated selection criteria.  The contours outline regions with visual extinctions greater than 8 mag.  The t'-t difference map at 5.0 Myr is shown in the bottom left corner.  
}
\end{center}
\end{figure*}

\begin{figure*}
\begin{center}
   \includegraphics[scale=0.63,clip=true,trim=1cm 0cm 0cm 0cm]{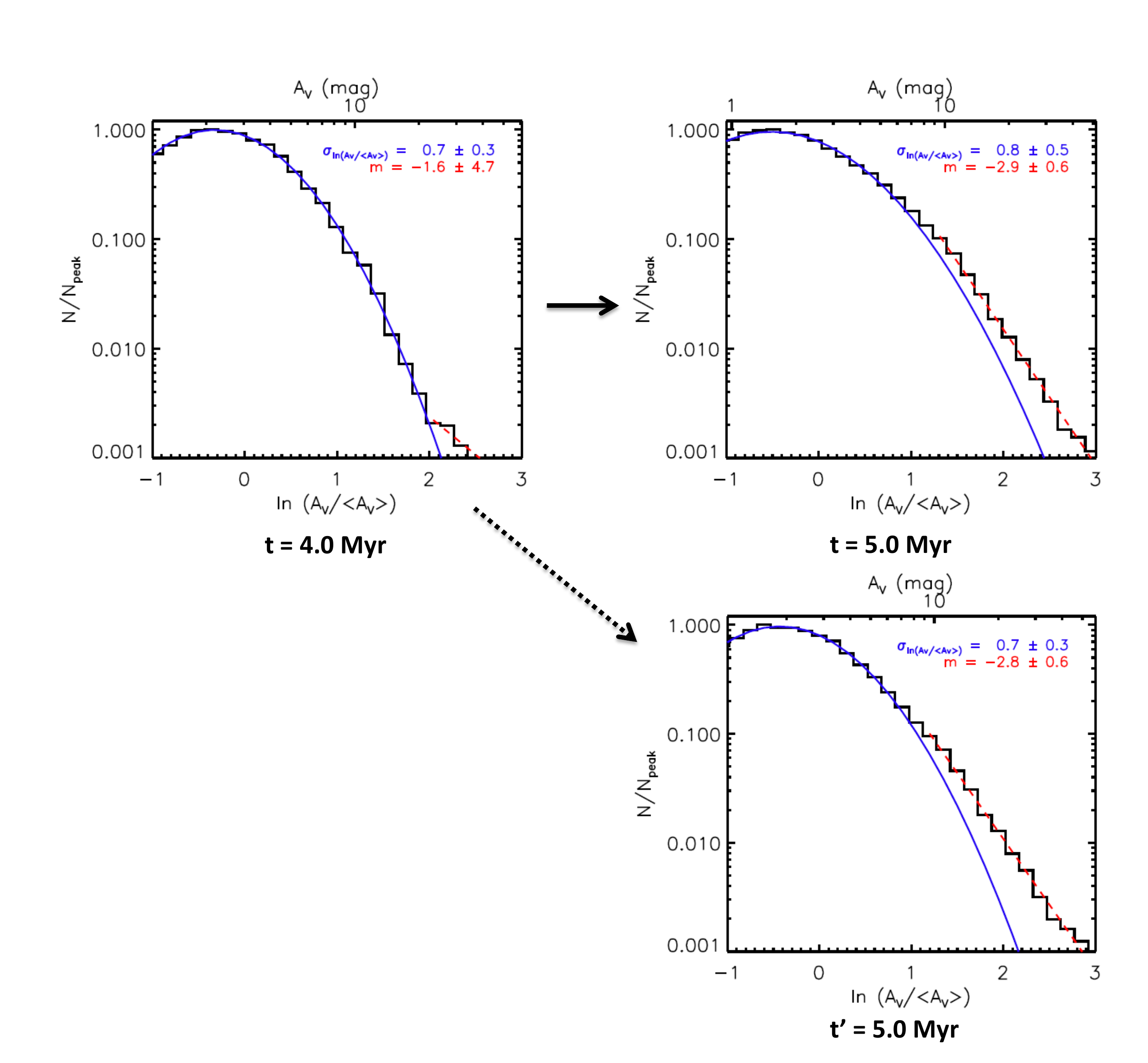} 
   \caption{\label{fg:evolclouds_pdfs}
  Column density PDFs for the evolution of a single cloud.  The column density PDFs are normalized by the peak of the distribution and are plotted as a function of the mean-normalized extinction, $\ln(A_{V}/\langle A_{V}\rangle)$. The solid line shows the lognormal fit to the distribution and the dashed line shows the power-law fit to the tail of the distribution.
}
\end{center}
\end{figure*}

To demonstrate that we can sensibly track the evolution of our clouds, a bound 10$^4$ M$_{\odot}$ cloud selected at t = 4.0 Myr is shown in Figures~\ref{fg:evolclouds_maps} and~\ref{fg:evolclouds_pdfs} to evolve in two possible ways.  For the true evolution (shown by the solid arrow), gas particles present in a column density map at t = 4.0 Myr are marked using their particle ids from the simulation, extracted from the simulation volume, and evolved forward in time.  This allows us to study the dynamics of the particles, as they are isolated from the surrounding medium.  For the perceived evolution (shown by the dotted arrow), we take two independently selected clouds: one in the 4.0 Myr snapshot and one in the 5.0 Myr snapshot.  If two clouds selected in this way have x-y positions in the map which are close, we claim that the older cloud is the evolved version of the younger cloud.  This method based on appearance is analogous to what observers do to better understand cloud evolution.  We note from Figure~\ref{fg:evolclouds_maps} that the two outcomes are quite alike in appearance; however, they are also similar in their bulk properties, such as their maximum surface densities and in the amount of mass in dense gas.  We also find that both methods of tracking the evolution of the cloud result in column density PDFs that are in agreement within uncertainties as seen in Figure~\ref{fg:evolclouds_pdfs}.  Therefore, our method of tracking cloud evolution is robust as the effects of the surrounding environment are minor and do not significantly influence the structural evolution of the cloud during the timescales of interest (on the order of a free-fall time, t$_{ff}$).  We will explore the effect of environment on longer timescales in future work for clouds formed in a galactic disc.  

\begin{figure}
 \begin{center}
  \includegraphics[scale=0.35, angle=0]{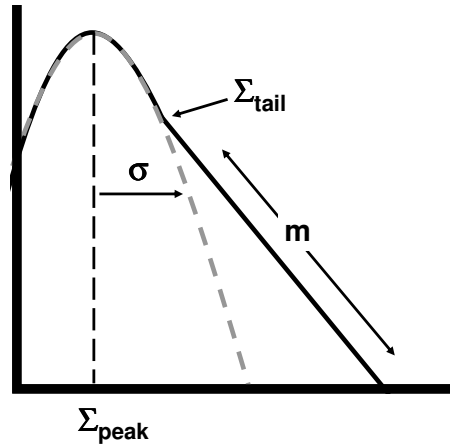}
  \caption{\label{fg:Npdf}
  Schematic diagram of the mass surface density probability distribution function (PDF).  The structural properties of interest are the peak, $\Sigma_{\text{peak}}$, and width of the distribution, $\sigma$, the slope of the power-law tail, m, and the surface density at which the power-law tail first forms, $\Sigma_{\text{tail}}$.  
}
 \end{center}
\end{figure}

We follow the evolution of each of our $\sim$ 60 clouds forward in time, producing a column density map after each 1 Myr time interval.  For each column density map, we produced a corresponding column density PDF plotted over a range chosen to match the observations of \citet{kain2009}.  To study the evolution of the molecular cloud structure in detail, we focus on the defining characteristics of the column density PDF.  The shape of the PDF is characterised by four key structural properties: its width ($\sigma$), peak location ($\Sigma_{peak}$), the slope of the power-law (m), and the deviation point ($\Sigma_{tail}$).  These properties are identified in the diagram shown in Figure~\ref{fg:Npdf}.

The peak and width of the distribution are determined by fitting the lognormal function 
\begin{equation}
p(\zeta) = \frac{1}{\sqrt{2\pi\sigma_{\zeta}^2}} \exp{\left[-\frac{(\zeta-\mu)^2}{2\sigma_{\zeta}^2}\right]}
\end{equation}
to the low extinction range of the PDF between $\zeta = \ln(A_{V}/\langle A_{V}\rangle)$ = [-1,1] \citep{kain2009,schneider2014} where $\mu$ and $\sigma_{\zeta}$ are the mean logarithmic column density and dispersion respectively.  The deviation point, $\Sigma_{\text{tail}}$, is defined as the value of $\zeta$ where the data deviates from the lognormal fit by a factor of 1.25 (dashed lines in Figure~\ref{fg:evolclouds_pdfs}).  This definition of the deviation point coincides with the values which would be identified by eye.  The power-law slopes are fit between the deviation point and the surface density at which $N/N_{peak}$ = 0.001, below which the signal would be too low to be detected observationally.  

\begin{figure*}
 \begin{center}
  \includegraphics[scale=0.6, angle=0, clip=true, trim=0 5mm 0 5mm]{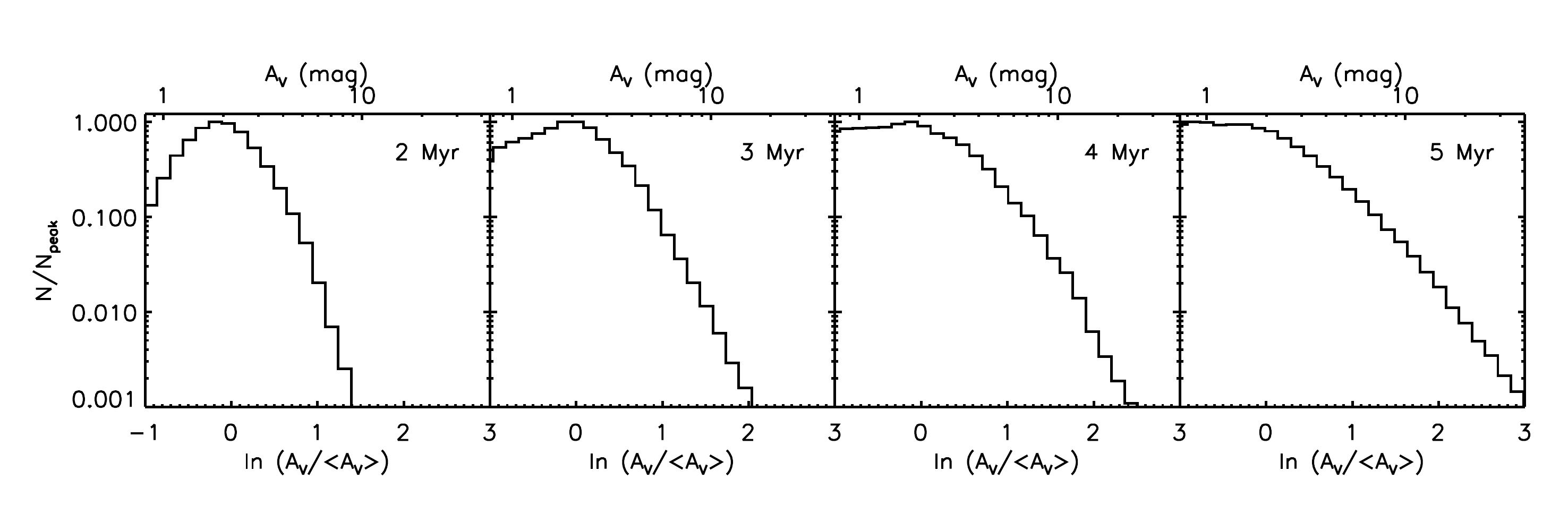}
  \caption{\label{fg:evolpdf}
  Evolution of the column density PDF for a typical molecular cloud.  Each PDF is normalized by the peak of its distribution and plotted as a function of the mean-normalized extinction, $\ln(A_{V}/\langle A_{V}\rangle)$. 
}
 \end{center}
\end{figure*}

\begin{figure*}
 \begin{center}
  \includegraphics[scale=0.8, angle=0, clip=true, trim= 0 3mm 0 7mm]{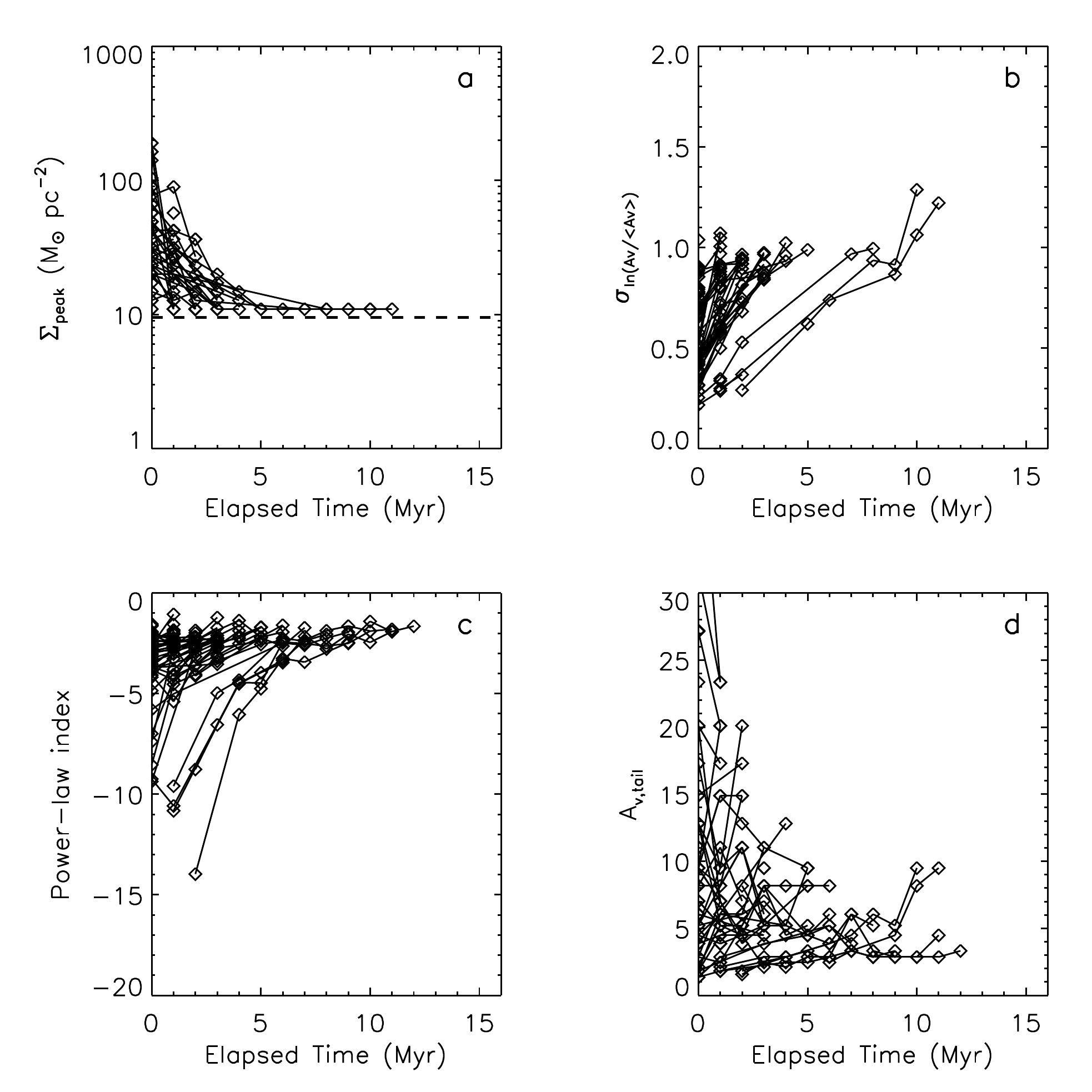}
  \caption{\label{fg:evols}
  Evolution of the structural properties of molecular cloud surface density PDFs: a) Peak surface density ($\Sigma_{\text{peak}}$), b) width ($\sigma_{\text{ln(Av/$\langle$Av$\rangle$)}}$), c) power-law slope (m), and d) tail deviation point ($A_{\text{V,tail}}$) as a function of time.  The solid lines show the evolutionary tracks for each cloud.  The dashed line shows the minimum surface density corresponding to the threshold for detection in observations.  
}
 \end{center}
\end{figure*}

At early times (t $\sim$ 0.1 t$_{ff}$), we find that the clouds appear diffuse and turbulent, exhibiting a lognormal distribution. At later times after the sinks have begun to form (t $\sim$ 0.6 t$_{ff}$ ), the distributions have developed a power-law tail along with an underlying lognormal shape, which is significantly wider with a lower peak column density (Figure~\ref{fg:evolpdf}).  To illustrate the characteristics of the structural transition and the times at which these transitions occur, we plotted the evolution of the four column density PDF properties for our clouds in Figure~\ref{fg:evols}.  

Clouds which could not be fit by a lognormal due to evolution of the peak to extinction values below the minimum threshold were excluded from Figures~\ref{fg:evols}(a) and \ref{fg:evols}(b) and are not considered in subsequent sections.  Since it was still possible to fit power-law slopes to these clouds, they are included in Figures~\ref{fg:evols}(c) and \ref{fg:evols}(d) and are considered in later discussions.  

In Figure~\ref{fg:evols}(a), we see that the peak of the distribution decreases as a function of time.  We see this trend for both bound and unbound clouds.  However, the peak of the distribution cannot evolve to surface densities any lower than the imposed threshold.  Therefore, there may be a significant amount of mass in clouds which exists below the threshold for detection.  This indicates that observational limitations can greatly impact the number of molecular clouds identified at late stages of their evolution.  

\begin{figure} 
\begin{center}
\includegraphics[scale=0.6, angle=0, clip=true, trim=0.5cm 0cm 0cm 0cm]{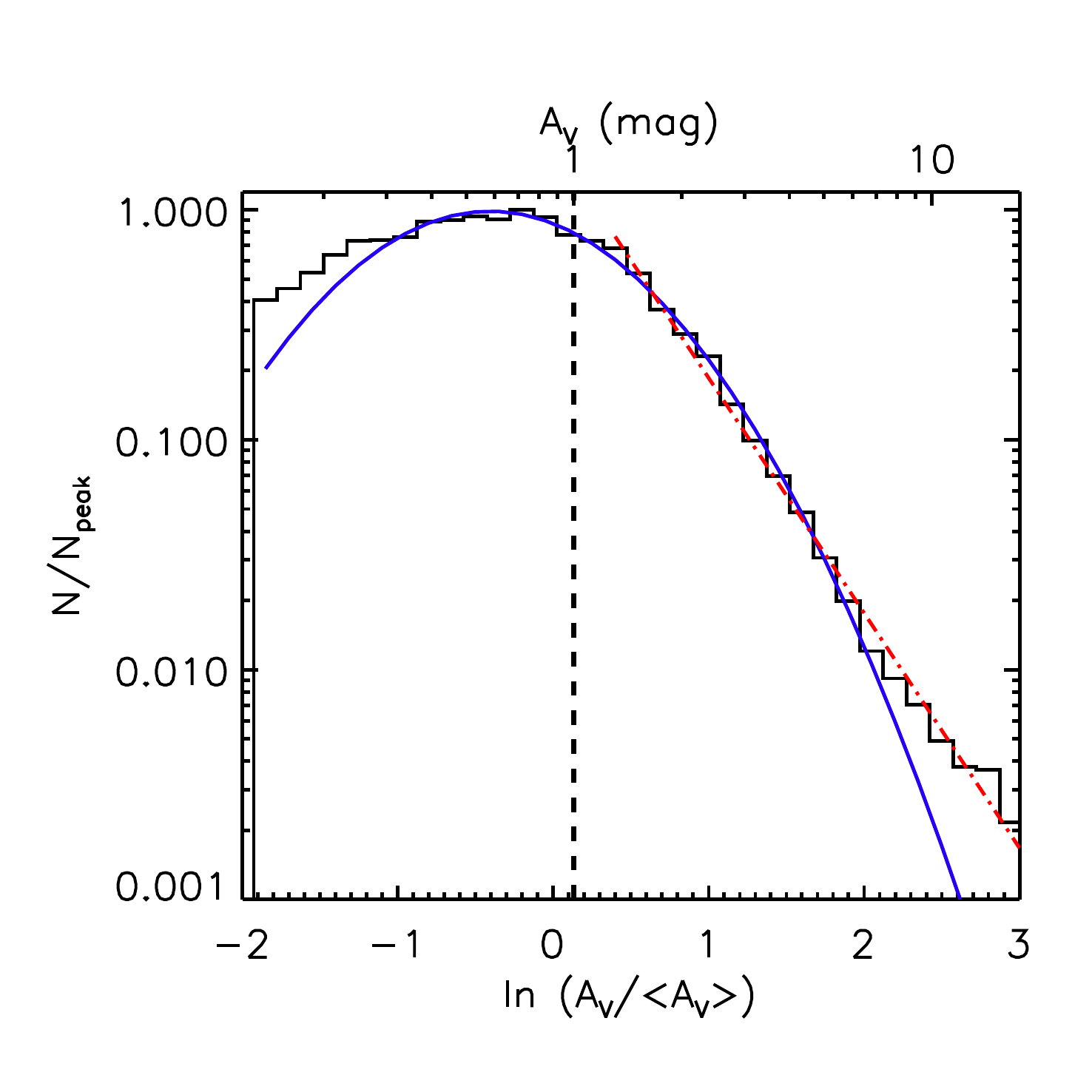} 
\caption{\label{fg:thresh}
 Column density PDF for a sample cloud at a late stage of its evolution ($t$ $>$  $t_{ff}$).  The solid line shows the lognormal fit to the distribution and the dash-dotted line shows the power-law fit to the tail of the distribution.  The dashed vertical line at $A_{\text{V}}$ = 1 mag represents the lower limit for detection adopted by \citet{lombardi2014}.
}
\end{center}
\end{figure}

\citet{lombardi2014} find no evidence of lognormality in their PDFs of the Orion star-forming region using data from Herschel and Planck.  The authors suggest that the lognormal could be confined to low column densities ($<$ A$_V$ $\sim$ 1 mag) below what they are able to detect and argue that the power-law regime dominates and characterises the majority of cloud structure.  We show that this is indeed the case for `old' clouds -- clouds which have been actively star-forming for several Myrs, like Orion.  Although stars are consuming gas, the amount of dense gas (n $>$ 10$^4$ cm$^{-3}$) available for star formation tends to remain relatively constant over time due to accretion and dispersal.  These clouds have begun dispersing the low column density gas and are rapidly converting high column density gas into stars, the net result of which is a lowering of the overall peak column density of the lognormal distribution.  The younger (or less evolved) a cloud is, the more likely a lognormal distribution will be observed in its PDF.  Alternatively, older star-forming regions will be dominated solely by a power-law as their distributions have evolved to peak below the current limits of detection by observations.   Figure~\ref{fg:thresh} shows the PDF for a cloud in our sample which would be identified as having a power-law across the entire range of column densities within the observed limits ($A_{\text{V}}$ $>$ 1 mag); however, we can see that the lognormal is still present, but now shifted below the lower limit for detection (dashed vertical line).

\begin{figure}
\begin{center}
\includegraphics[scale=0.7, angle=0, clip=true, trim=0.5cm 0cm 0cm 0cm]{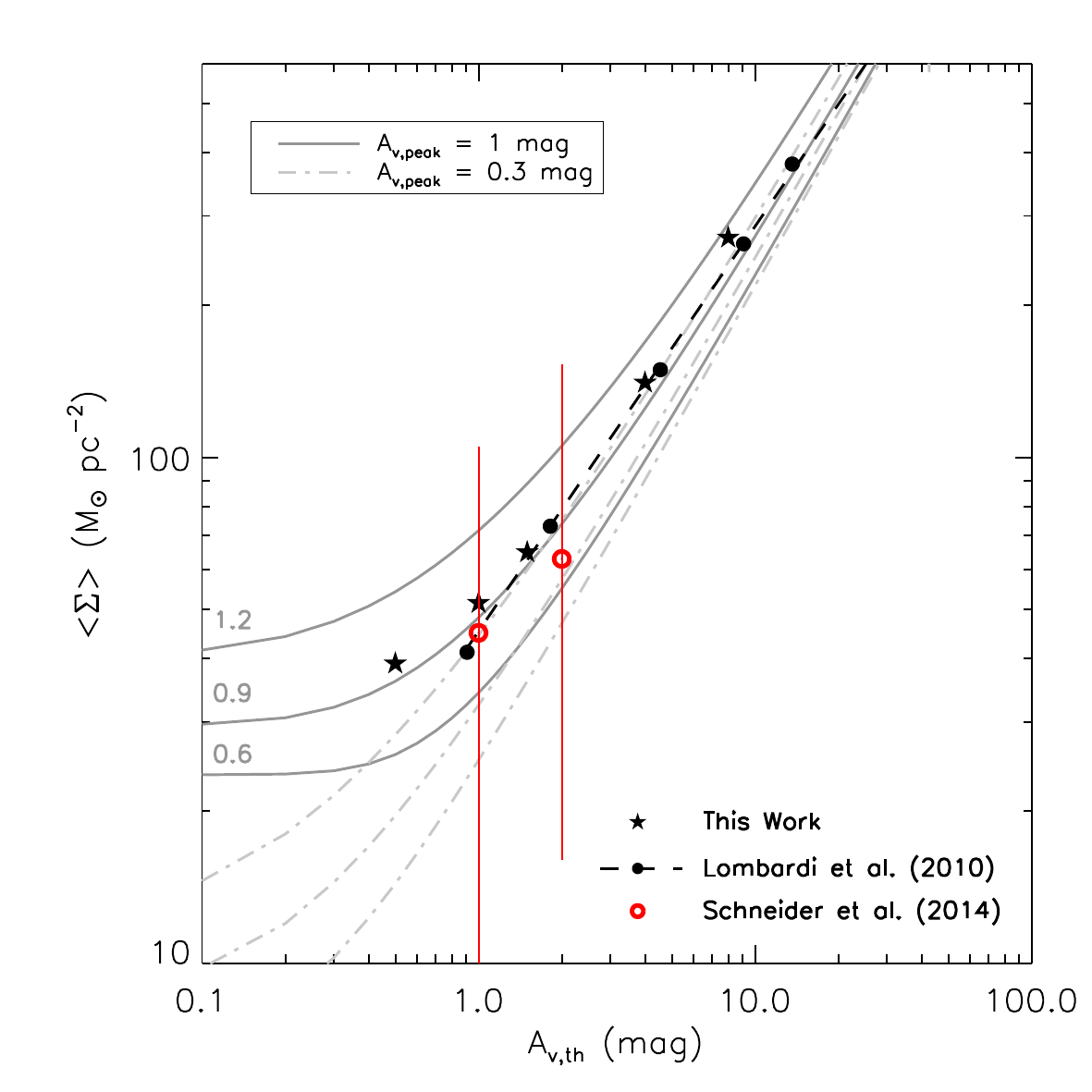}
\caption{\label{fg:sigAth}
 Mean mass surface density as a function of the threshold, given by the corresponding extinction threshold, A$_{\text{v,th}}$.  The black stars represent the mean mass surface densities of our sample of simulated molecular clouds for five thresholds corresponding to A$_{\text{v,th}}$ = [0.5, 1, 1.5, 4, 8].  The black dots represent the mean mass surface densities for a sample of \emph{observed} clouds obtained from Table 1 of \citet{lombardi2010}.  The black dashed line is a power-law fit to the observed data from \citet{lombardi2010}.  The grey lines are analytic calculations of the mean surface density as a function of the threshold assuming a lognormal column density probability distribution function (PDF) with a peak at A$_{\text{v,peak}}$ = 1 mag (solid line) and at A$_{\text{v,peak}}$ = 0.3 mag (dash-dotted line) and standard deviations of $\sigma$ = [0.6, 0.9, 1.2] \citep{lombardi2010, balle2012}. 
}
\end{center}
\end{figure}

We explore the effect of a threshold further by measuring the mean mass surface densities, $\langle\Sigma\rangle$, for our sample of simulated clouds assuming five possible extinction thresholds (A$_{\text{v,th}}$ (mag) = 0.5, 1.0, 1.5, 4, 8) and comparing them to the mean mass surface densities of observed molecular clouds. Only pixels with column densities above a given threshold contribute to estimates of $\langle\Sigma\rangle$. Using near-IR extinction observations of molecular clouds, \citet{lombardi2010} measured $\langle\Sigma\rangle$ for five different K-band extinction thresholds: A$_{\text{K,th}}$ = [0.1, 0.2, 0.5, 1.0, 1.5] \citep[$A_{\text{K}}/A_{\text{V}}$ = 0.11;][]{RL85} and found that different clouds will have a constant column density for a given threshold and the value of the column density is dependent on the threshold. Figure~\ref{fg:sigAth} shows the agreement between our results (black stars) and the observational estimates of $\langle\Sigma\rangle$ by \citet{lombardi2010} (black dots) along with their corresponding fit, $\Sigma = 265 \text{ M}_{\odot} \text{ pc}^{-2} (A_{th}/\text{mag})^{0.8}$ (black dashed line).  The most recent estimates for mean surface densities of nearby clouds from Schneider et al. (2014) are also shown (red circles).
 
\citet{lombardi2010} argue that their results are a consequence of cloud structure described by a lognormal column density PDF.  However, \citet{beaumont2012} and \citet{balle2012} both suggest that alternate forms of the column density PDF could also produce these results provided that the chosen thresholds are near or above the peak of the distribution.  In particular, \citet{balle2012} explored the consequences of several different functional forms of the column density PDF.  While they argue that the results of \citet{lombardi2010} are an effect of thresholding the surface density, their analytic models do show that a lognormal column density PDF with or without a power-law tail can fit the observations.  In Figure~\ref{fg:sigAth}, we plotted six possible cases of their model for a lognormal-shaped column density PDF, shown as grey lines, where

\begin{equation}
\begin{split}
   \langle\Sigma(\text{A}_{\text{v,th}})\rangle = (17.5 \text{M}_{\odot} \text{pc}^{-2})\text{A}_{\text{peak}}\exp(\frac{\sigma^2}{2}) \\
   \times \frac{1-\text{erf}\{[\text{ln}(\text{A}_{\text{v,th}}/\text{A}_{\text{peak}})-\sigma^2]/\sqrt{2}\sigma\}}{1-\text{erf}[\text{ln}(\text{A}_{\text{v,th}}/\text{A}_{\text{peak}})/\sqrt{2}\sigma]}
   \label{eq:1}
\end{split}
\end{equation}

\noindent \citep{balle2012} with A$_{\text{peak}}$ = 1 mag (shown as solid lines) or A$_{\text{peak}}$ = 0.3 mag (shown as dash-dotted lines) and $\sigma$ = 0.6, 0.9, and 1.2 for the widths of each A$_{\text{peak}}$ distribution.  We find that a lognormal distribution is consistent with both the observational data of \citet{lombardi2010} and our simulated clouds.  The analytic solutions plateau for extinction thresholds less than the peak for the A$_{\text{peak}}$ = 1 mag case.  However, once the peak of the distribution evolves to lower surface densities (and correspondingly lower extinctions) such as for the A$_{\text{peak}}$ = 0.3 mag case, the mean surface density is always a function of the threshold across the entire range of threshold values. 

We have shown that due to the peaked distribution of the column density PDF, great care must be taken when determining the mean mass surface density as the estimate depends sensitively on the choice of threshold.  Clouds disperse as they evolve, resulting in localised high density regions within a cloud dominated by low-density molecular gas.  Provided that the threshold is below or near the peak of the distribution, the measure of mean cloud mass surface density can be trusted; however at late times, many of the cloud properties, including the mass and area, will be largely underestimated as they now have evolved to become a function of the threshold.  

Cloud dispersal can also lead to a wider lognormal surface density PDF.  This widening, seen in Figure~\ref{fg:evols}(b), is expected for clouds with increasing Mach numbers (see section~\ref{sec:discussion}) and is in agreement with \citet{balle2011} and \citet{tassis} who have shown that the distribution will widen prior to the development of a power-law tail.  

The power-law tail is an interesting feature of the column density PDF.  Many authors \citep[e.g.][]{knw11,tassis,balle2011} have shown that the development of a power-law tail occurs as the cloud evolves from being turbulence-dominated to being gravity-dominated.  However, there is still a question as to whether or not the slope of the tail is universal.  \citet{froebrich07} attribute a variation in the slope to cloud distance based on 2MASS observations of molecular clouds.  \citet{balle2011} also argue against a universal slope as they find variation due to projection in their simulations of clouds formed at the interface of colliding streams.  However, neither of these interpretations account for cloud age and evolution.

In Figure~\ref{fg:evols}(c), we show the slopes of the power-law tails obtained from the column density PDFs of our simulated clouds as a function of time.  At early times, the slopes are very steep, as they coincide with the lognormal fits at high column densities.  As the clouds evolve, the slopes become more shallow, eventually converging to a power-law slope of approximately -2.   This result is in agreement with \citet{knw11} who also do not see large variation in their power-law slopes with a range of values between -2.8 and -2.  Previous studies which have found power-law tails with varying slopes are consistent with our findings for clouds at early stages of their evolution, but our results show that the slopes trend towards a constant value at late times.  A universal slope for the high-mass end of molecular cloud structure is appealing as it links the column density distribution of the gas to the initial mass function (IMF) of stars.  However, this slope is largely dependent on the point, $\Sigma_{\text{tail}}$, where the tail begins its divergence from the lognormal distribution.  Determining whether or not the deviation point is constant will aid in our understanding of the universality of the slope and its relation to star formation.

Figure~\ref{fg:evols}(d) shows the evolution of the deviation point as a function of time.  We find a decreasing trend towards constant values of the deviation point, consistent with observations \citep{kain2011,schneider2014}.  The reason for this is that the peak of the distribution is also changing with time (Figure~\ref{fg:evols}(a)).  As the distribution widens, the peak shifts to lower extinction values, resulting in a relatively constant transition point at $A_{\text{V,tail}}$ $\sim$ 4 mag for the power-law tail at late times.  This deviation point is consistent with the value estimated from 2MASS observations by \citet{kain2011} (A$_{\text{v}}^{\text{tail}}$ $\approx$ 2 -- 4 mag) and from Herschel observations by \citet{schneider2014} (A$_{\text{v}}^{\text{tail}}$ $\approx$ 4 -- 5 mag).  Once gravity begins to dominate, the tail deviates from the lognormal and grows to have a shallower slope.  The peak of the distribution is simultaneously decreasing so, by the time star formation takes hold, the dense gas component is entirely confined to the tail of the distribution for surface density values greater than a constant deviation point of $A_{\text{V,tail}}$ $>$ 4 mag. 

\begin{figure}
 \begin{center}
  \includegraphics[scale=0.7, angle=0]{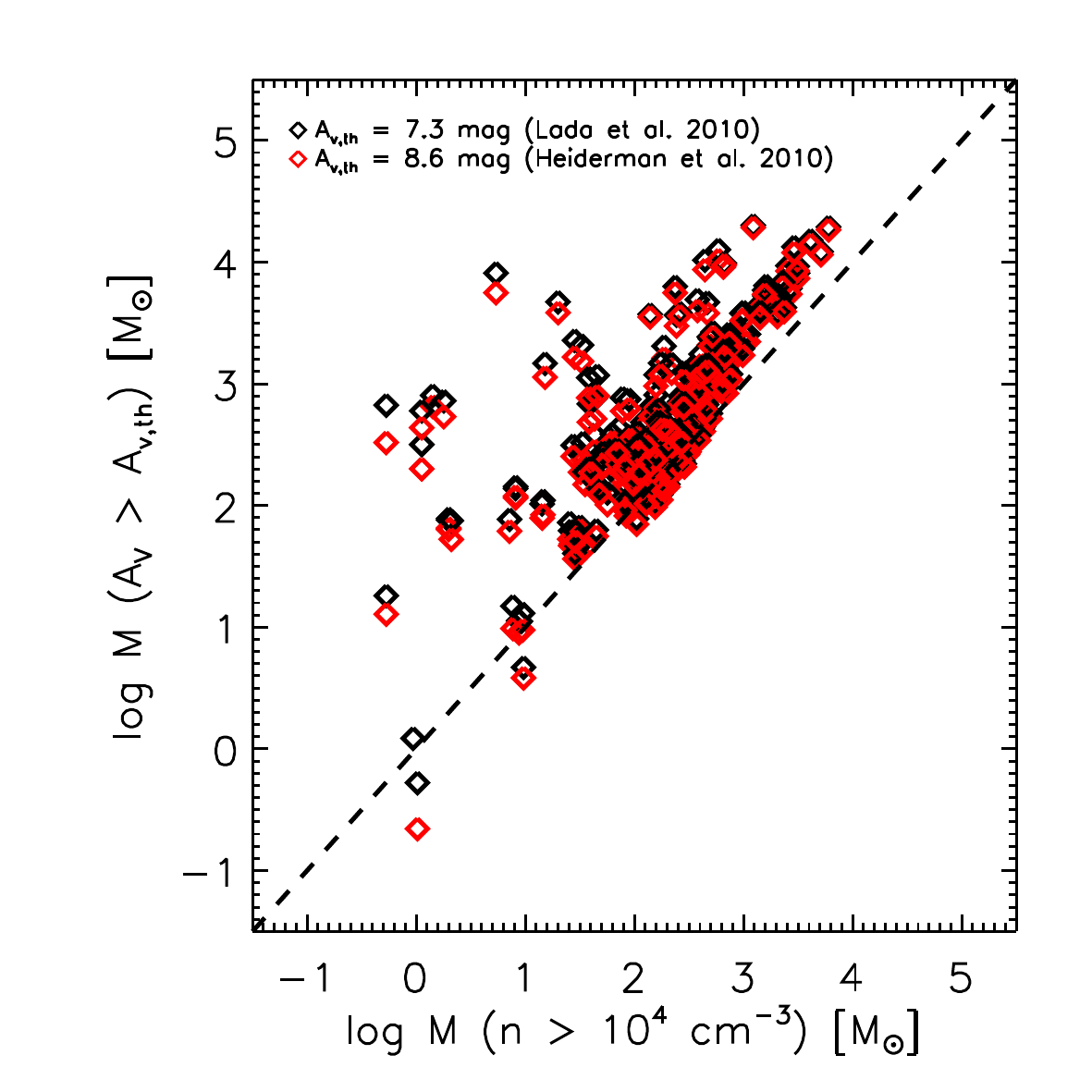}
  \caption{\label{fg:sfgas}
  Mass found in dense gas (n $>$ 10$^4$ cm$^{-3}$) compared to the mass found above an extinction of $A_{\text{V}}$ = 7.3 mag (black diamonds) or A$_{\text{V}}$ = 8.6 mag (red diamonds).  The one-to-one line is also shown as a black dashed line.
}
 \end{center}
\end{figure}

\citet{lada2010} define an extinction threshold for star formation, above which the surface density of the participating gas correlates linearly with the star formation rate.  The authors assume that the material at visual extinctions of greater than 7.3 mag corresponds to a gas volume density of 10$^4$ cm$^{-3}$ \citep{bergin01}.  \citet{heiderman2010} also found the presence of a star formation threshold in an independent study of young stellar objects (YSOs) using the Spitzer c2d and Gould Belt surveys; however, this threshold was at a higher visual extinction of 8.6 mag.  Both of these estimates of the star formation threshold are greater than the value of the deviation point found in this work and by other authors \citep{kain2011,schneider2014}; however, $A_{\text{V}}^{\text{tail}}$ by its construction is dominated by the diffuse component of the underlying lognormal PDF.  \citet{kain2011} argue that the surface density where the contribution from the tail dominates more than 90\% of the PDF is a more direct comparison to the star formation thresholds.  Since we know the volume density of the gas particles in our simulation, we can determine where the dense component of the gas is dominant in our clouds.  In Figure~\ref{fg:sfgas}, we compare the mass in high volume density regions to the mass found at high extinction in our own synthetic maps for the two estimates of the star formation threshold. 

This figure shows that masses derived from column density maps will be largely overestimated if one assumes that the star formation threshold is an effective tracer of the dense gas regions corresponding to active star formation.  We note however, that while in both cases we find that the mass in high extinction gas is much greater than the mass in high density gas, there still appears to be a correlation, particularly for masses greater than 10$^3$ M$_{\odot}$.  As the clouds evolve, the mass in dense gas increases and eventually collapses to form stars.  This would explain why the correlation is much tighter at higher masses, as it corresponds to later stages of cloud evolution when the cloud begins to actively form stars.  This correlation also suggests that we should expect a similar relation as that found by \citet{lada2010}, where the star formation rate (SFR) scales as a function of the mass of high extinction material such that,
\begin{equation}
	SFR \text{ (M}_{\odot} \text{ yr}^{-1}) = 4.6 \pm 2.6 \times 10^{-8} M_{\text{A$_\text{V}$} > 7.3} \text{ (M}_{\odot})
\end{equation}

\begin{figure}
 \begin{center}
  \includegraphics[scale=0.7, angle=0]{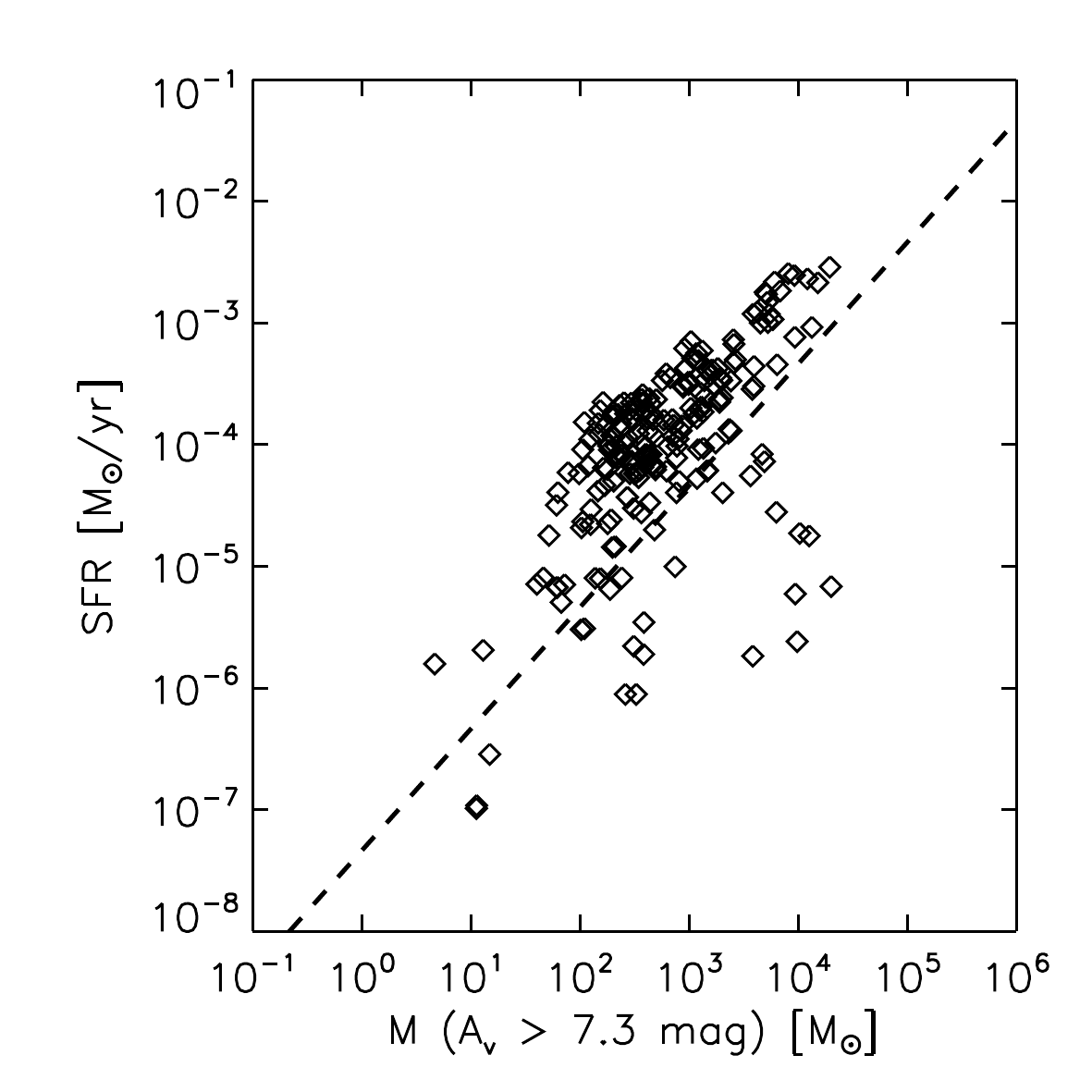}
  \caption{\label{fg:sfrs}
Star formation rate as a function of mass above an extinction threshold of $A_{\text{V}}$ = 7.3 mag.  The best fit to the observations found by \citet{lada2010} is also shown as a dashed line.
  }
 \end{center}
\end{figure}

We calculated the star formation rate for our simulated clouds by dividing the total mass in stars at each output by the total time elapsed since the beginning of the simulation.  Figure~\ref{fg:sfrs} shows the star formation rate in our clouds as a function of the cloud mass above the extinction threshold $A_{\text{V}}$ = 7.3 mag.

We find that the star formation rate of molecular clouds correlates very well with the mass above a visual extinction of 7.3 mag.  We find the best fit to our data is $SFR$ = 10.7 $\times$ 10$^{-8}$ M$_{\text{A$_\text{V}$} > 7.3}$, where the coefficient is a factor of approximately 2 larger than that found by \citet{lada2010}.  This could be due to a higher star formation efficiency in our simulations than for the observed clouds.  Nevertheless, the existence of the correlation confirms the relation between the mass in dense gas and the mass at high extinction as well as the significance of a constant star formation threshold.  The evolution towards a constant deviation point is the result of the interplay between all other structural properties of the column density PDF.  A constant value for the deviation point at late times leads to a universal slope of the power-law tail: structural properties which are directly linked to the star formation threshold and slope of the high-mass end of the IMF, respectively. 

\section{Conclusions}\label{sec:discussion}

In this paper, we studied the evolution of the column density PDF of molecular clouds using simulations and synthetic column density maps.  For bound and unbound clouds which form from a variety of initial conditions, we find that the column density PDF is consistently represented by a lognormal distribution at early times (t $<$ 0.25 t$_{ff}$ ) with a structural transition to a lognormal + power-law tail at late times (t $>$ 0.5 t$_{ff}$ ).  The evolution of the four key structural properties ($\sigma$, $\Sigma_{peak}$, m, and $\Sigma_{tail}$) were explored in detail.  We found that, as a cloud evolves, the peak of its column density PDF will shift to lower extinction values as the distribution widens regardless of whether or not it is gravitationally bound.  This means that the underlying lognormal distribution, which can be present for a cloud at early times, will be lost for that same cloud at late times due to the evolution of the peak below the observational threshold for detection.  Our results help to explain why certain observations of actively star-forming, dynamically older clouds, such as the Orion molecular cloud, do not appear to have any evidence of a lognormal distribution in their column density PDFs and are instead represented by a power-law tail over the full range of extinction \citep{lombardi2014}.  We also find an increase in the column density variance, even though we do not have externally driven turbulence in our simulations.  Therefore, the widening of the distribution does not depend solely on the supersonic turbulence \citep[see also][]{tassis}, but arises as a result of density perturbations produced by localised gravitational collapse.  

By studying the evolution of the slope and deviation point of the power-law tails for our sample of simulated clouds, we also show that the range for both properties narrows as they trend towards constant values, thus linking the column density structure of the molecular cloud to such stellar properties as the initial mass function and the surface density threshold for star formation.  Although the mass in high extinction gas is much greater than the mass in dense gas, we still find a correlation between the two, particularly at masses greater than 10$^3$ M$_{\odot}$.  We also show that the thresholds used to define the minimum surface density of gas participating in star formation \citep{lada2010,heiderman2010} are suitable tracers of the star formation rates in molecular clouds.

A study of the statistics of cloud properties for a large population of molecular clouds formed in a galactic disc will be the subject of future work.  This study will allow for a comparison of our results to clouds observed in galactic and extragalactic environments, as well as further explore the implications of a constant star formation threshold on the star formation rates of bound and unbound molecular clouds.  

\section*{Acknowledgments}

We would like to thank SHARCNET (Shared Hierarchical Academic Research Computing Network) 
and Compute/Calcul Canada, which provided dedicated resources to run these simulations.   
This work was supported by NSERC.  J.W. acknowledges support from the Ontario Early Researcher Award (ERA).

\bibliographystyle{apj}

\begin{thebibliography}{}

\bibitem[{{Ballesteros-Paredes} et al.(2011)}]{balle2011}
{Ballesteros-Paredes}, J., {V\'azquez-Semadeni}, E., {Gazol}, A., et al. \ 2011, MNRAS, 416, 1436

\bibitem[{{Ballesteros-Paredes} et al.(2012)}]{balle2012}
{Ballesteros-Paredes}, J., {D'Alessio}, P., \& {Hartmann}, W.~L. \ 2012, MNRAS, 427, 2562

\bibitem[{{Bate} \& {Bonnell}(2005)}]{BB05}
{Bate}, M.~R. \& {Bonnell}, I.~A. 2005, MNRAS, 356, 1201 

\bibitem[{{Beaumont} et al.(2012)}]{beaumont2012}
{Beaumont}, C.~N., {Goodman}, A.~A., {Alves}, J.~F., et al. \ 2012, MNRAS, 423, 2579

\bibitem[{{Bergin} et al.(2001)}]{bergin01}
{Bergin}, E.~A., {Ciardi}, D.~R., {Lada}, C.~J., {Alves}, J., \& {Lada}, E.~A. \ 2001, ApJ, 557, 209

\bibitem[{{Bohlin} et al.(1978)}]{bohlin}
{Bohlin}, R.~C., {Savage}, B.~D., \& {Drake}, J.~F. \ 1978, ApJ, 224, 132

\bibitem[{{Chabrier}(2003)}]{chabrier}
{Chabrier}, G. \ 2003, PASP, 115, 763

\bibitem[{{Federrath} et al.(2010)}]{federrath10}
{Federrath}, C., {Banerjee}, R., {Clark}, P.~C., \& {Klessen}, R.~S. \ 2010, ApJ, 713, 269 

\bibitem[{{Froebrich} et al.(2007)}]{froebrich07}
{Froebrich}, D., {Murphy}, G~C., {Smith}, M.~D., {Walsh}, J. \& {del' Burgo}, C. \ 2007, MNRAS, 378, 1447

\bibitem[{{Goodman} et al.(2009)}]{goodman09}
{Goodman}, A.~A., {Pineda}, J.~E., \& {Schnee}, S.~L. \ 2009, ApJ, 692, 91

\bibitem[{{Heiderman} et al.(2010)}]{heiderman2010}
{Heiderman}, A., {Evans}, N.~J., {Allen}, L.~E., {Huard}, T., \& {Heyer}, M. \ 2010, ApJ, 723, 1019

\bibitem[{{Hennebelle} \& {Chabrier}(2008)}]{HC08}
{Hennebelle}, P. \& {Chabrier}, G. \ 2008, ApJ, 684, 395

\bibitem[{{Kainulainen} et al.(2009)}]{kain2009}
{Kainulainen}, J., {Lada}, C.~J., {Rathborne}, J.~M., \& {Alves}, J.~F. \ 2009, A\&A, 497, 399

\bibitem[{{Kainulainen} et al.(2011)}]{kain2011}
{Kainulainen}, J., {Beuther}, H., {Banerjee}, R., {Federrath}, C., \& {Henning}, T. \ 2011, A\&A, 530, A64

\bibitem[{Kainulainen}, {Federrath}, \& {Henning}(2013)]{KFH13}
{Kainulainen}, J., {Federrath}, C., \& {Henning}, T. \ 2013, A\&A, 553, 8

\bibitem[{{Kritsuk} et al.(2011)}]{knw11}
{Kritsuk}, A.~G., {Norman}, M.~L., \& {Wagner}, R. \ 2011, ApJ, 727, L20

\bibitem[{{Kroupa}(2001)}]{kroupa}
{Kroupa}, P. \ 2001, MNRAS, 322, 231

\bibitem[{{Krumholz} \& {McKee}(2005)}]{KM05}
{Krumholz}, M.~R. \& {McKee}, C.~F. \ 2005, ApJ, 630, 250 

\bibitem[{{Lada} et al.(2010)}]{lada2010}
{Lada}, C.~J., {Lombardi}, M., \& {Alves}, J.~F. \ 2010, ApJ, 724, 687

\bibitem[{{Lombardi} et al.(2010)}]{lombardi2010}
{Lombardi}, M., {Alves}, J., \& {Lada}, C. J. \ 2010, A\&A, 519, 7

\bibitem[{{Lombardi} et al.(2014)}]{lombardi2014}
{Lombardi}, M., {Bouy}, H., {Alves}, J., \& {Lada}, C. J. \ 2014, A\&A, 566, A45

\bibitem[{{Nordlund} \& {Padoan}(1999)}]{NP99}
{Nordlund}, \AA. \& {Padoan}, P. \ 1999, Interstellar Turbulence, ed. J. Franco \& A. Carraminana, 218

\bibitem[{Nutter} et al.(2008)]{nutter}
{Nutter}, D., {Kirk}, J. M., {Stamatellos}, D., \& {Ward-Thompson}, D. \ 2008, MNRAS, 384, 755

\bibitem[{{Ostriker} et al.}(2001)]{ostriker2001}
{Ostriker}, E.~C., {Stone}, J.~M., \& {Gammie}, C.~F. \ 2001, ApJ, 546, 980

\bibitem[{{Padoan} \& {Nordlund}(2002)}]{PN02}
{Padoan}, P. \& {Nordlund}, \AA. \ 2002, ApJ, 576, 870

\bibitem[{{Rieke} \& {Lebofsky}(1985)}]{RL85}
{Rieke}, G.~H. \& {Lebofsky}, M.~J. \ 1985, ApJ, 288, 618

\bibitem[{{Salpeter}(1955)}]{salpeter}
{Salpeter}, E.~E. \ 1955, ApJ, 121, 161

\bibitem[{{Schneider} et al.(2014)}]{schneider2014}
{Schneider}, N., {Ossenkopf}, V., {Csengeri}, T., et al. \ 2014, A\&A, submitted

\bibitem[{{Tassis} et al.(2010)}]{tassis}
{Tassis}, K., {Christie}, D.~A., {Urban}, A., et al. \ 2010, MNRAS, 408, 1089 

\bibitem[{{V\'azquez-Semadeni}(1994)}]{vazquez1994}
{V\'azquez-Semadeni}, E. \ 1994, ApJ, 423, 681

\bibitem[{{Wadsley} et al.(2004)}]{gasoline}
{Wadsley}, J.~W., {Stadel}, J., \& {Quinn}, T. 2004, New Astronomy, 9, 137 

\bibitem[{{Ward} et al.(2012)}]{ward2012}
{Ward}, R.~L., {Wadsley}, J., {Sills}, A., \& {Petitclerc}, N. \ 2012, ApJ, 756, 119

\bibitem[{{Ward} et al.(2014)}]{paper1}
{Ward}, R.~L., {Wadsley}, J., \& {Sills}, A. \ 2014, MNRAS, 439, 651

\end{thebibliography}

\end{document}